\documentclass [aps,pra,amssymb,amsmath,twocolumn,showpacs]{revtex4-1}

\usepackage{amsmath}
\usepackage{amssymb}
\usepackage{graphicx}
\usepackage{verbatim}
\usepackage{bm}
\usepackage[dvipsnames]{xcolor}

\usepackage{graphicx,epsfig,amsfonts,amssymb}
\usepackage{hyperref}
  \hypersetup{colorlinks=true,citecolor=blue}

\usepackage{bm}
\usepackage{times}
\usepackage{lipsum}

\usepackage[all]{xy}

\newcommand{\nn}{\nonumber}

\newcommand{\la}{\langle}
\newcommand{\ra}{\rangle}
\newcommand{\rar}{\rightarrow}

\newcommand{\be}{\begin{eqnarray}}
\newcommand{\ee}{\end{eqnarray}}

\newcommand{\hc}{\hat{c}}
\newcommand{\hd}{\hat d}
\newcommand{\dg}{\dagger}
\newcommand{\hn}{\hat n}
\newcommand{\bs}{\begin{equation}\begin{split}}
\newcommand{\es}{\end{split}\end{equation}}

\date{\today}

\begin{document}
\title{Integrable multistate Landau-Zener models with parallel energy levels}

\author{Vladimir Y. Chernyak$^{a, b}$, Fuxiang Li$^c$, Chen Sun$^{d}$, and Nikolai A. Sinitsyn$^e$}
\affiliation{$^a$Department of Chemistry, Wayne State University, 5101 Cass Ave, Detroit, Michigan 48202, USA}
\affiliation{$^b$Department of Mathematics, Wayne State University, 656 W. Kirby, Detroit, Michigan 48202, USA}
\affiliation{$^c$School of Physics and Electronics, Hunan University, Changsha 410082, China}
\affiliation{$^d$Department of Physics, Brown University, Providence, Rhode Island 02912, USA}
\affiliation{$^e$Theoretical Division, Los Alamos National Laboratory, Los Alamos, NM 87545, USA}

\begin{abstract}
We discuss  solvable multistate Landau-Zener (MLZ) models whose Hamiltonians have  commuting partner operators with  $\sim 1/\tau$-time-dependent parameters.  Many already known solvable MLZ models belong precisely to this class.  We derive the integrability conditions on the parameters of such commuting operators, and demonstrate how to use such conditions in order to derive new  solvable cases.
We show that MLZ models from this class must contain bands of parallel diabatic energy levels. The structure of the scattering matrix and other properties are found to be the same as in the previously discussed completely solvable MLZ Hamiltonians. 
\end{abstract}
\maketitle

\section{Introduction}
\label{sec:intro}

This article is a continuation of a series of our publications \cite{quest-LZ,large-class,multitime-LZ} about properties and classification of solvable MLZ models that describe evolution with linearly time-dependent Hamiltonians:
\begin{eqnarray}
\label{mlz} i \frac{\partial \Psi }{\partial t} = H(t) \Psi, \quad H(t)=B t + A, 
\end{eqnarray}
where $B$ and $A$ are real symmetric time-independent $N\times N$ matrices with constant parameters. The basis that diagonalizes $B$ is called the diabatic basis. The off-diagonal elements of $A$ in this basis are called coupling constants. 

Many solvable models of this type have been found recently due to the discovery of integrability conditions \cite{quest-LZ} that impose constraints on the parameters of $H(t)$ such that the scattering matrix for the evolution from $t\rar -\infty$ to $t\rar +\infty$ can be constructed explicitly. In terms of a more general theory of integrable time-dependent Hamiltonians \cite{commute} the model (\ref{mlz}) can be fully solvable if for its Hamiltonian we can find another nontrivial Hamiltonian $H'$ such that 
\begin{eqnarray}
\label{cond1}
&&\partial_{\tau}  H(t,\tau) =\partial_t H'(t,\tau), \\
\label{cond2}
&& [H(t,\tau), H'(t,\tau)]=0,
\end{eqnarray}
where $\tau$ is one of the parameters of $H$. By ``solvable", in the MLZ theory, we mean that we can write the scattering matrix of the model for  the evolution from $t=-\infty$ to $t=+\infty$ in terms of commonly known special functions of the model's parameters.
Often, only  the matrix of transition probabilities is needed. Its entries are the absolute square elements of the scattering matrix \cite{quest-LZ}. 

 Note that due to the symmetry in (\ref{cond1}) and (\ref{cond2}), if Eq.~(\ref{mlz}) with $H(t,\tau)$ is solvable then the Schr\"odinger equation is also likely solvable for the Hamiltonian $H'$, treating $t$ as a constant parameter and $\tau$ as time of this new model.   The theory in \cite{commute} is very general, e.g., it operates with families of many operators that satisfy conditions (\ref{cond1})-(\ref{cond2}) simultaneously. Hence, our goal here is to narrow it and find the minimal form of the operators $H$ and $H'$ that is sufficient to describe practically all known solvable MLZ systems.

Considering Eq.~(\ref{cond1}) for a fixed value of $\tau$ as an ordinary differential equation  with respect to $t$, we solve it, resulting in
\begin{eqnarray}
\label{zero-curv-nonabel-2D-3} H'(t, \tau)= \frac{1}{2} \partial_{\tau}B(\tau)t^{2} + \partial_{\tau}A(\tau)t + D(\tau),
\end{eqnarray}
where $B$ and $A$ matrices are the same as in (\ref{mlz}), and $D$ is a new $t$-independent matrix.

Thus, $ H'(t,\tau)$ is generally represented by a quadratic polynomial of $t$. 
Substituting further Eq.~(\ref{zero-curv-nonabel-2D-3}) into Eq.~(\ref{cond2}) we arrive at
\begin{eqnarray}
\label{zero-curv-nonabel-2D-5a} && [\partial_{\tau} B(\tau), B(\tau)] = 0, \\
\label{zero-curv-nonabel-2D-5b} && \frac{1}{2} [\partial_{\tau}B(\tau), A(\tau)] + [\partial_{\tau} A(\tau), B(\tau)] = 0, \\
\label{zero-curv-nonabel-2D-5c} && [\partial_{\tau} A(\tau), A(\tau)] + [D(\tau), B(\tau)] = 0, \\
\label{zero-curv-nonabel-2D-5d} && [D(\tau), A(\tau)] = 0.
\end{eqnarray}
Equations~(\ref{zero-curv-nonabel-2D-5a})-(\ref{zero-curv-nonabel-2D-5d}) determine all possible integrable MLZ problems.

\section{$t/\tau$-family}
In \cite{multitime-LZ},  we discussed a particular family of models for which these equations are resolved by Hamiltonians with only linear dependence on $t$ and $\tau$: 
\begin{eqnarray}
\label{linear}
\nonumber 
H&=&B_{00}t + B_{01}\tau +A_0,\\
H'&=&B_{11}\tau + B_{01} t +A_1,
\ee
where $B_{00}$, $B_{01}$, and $B_{01}$ are mutually commuting, and therefore  simultaneously diagonalizable, matrices. Equations~(\ref{zero-curv-nonabel-2D-5a})-(\ref{zero-curv-nonabel-2D-5d}) allow an extension of this class by assuming that
$$
D(\tau)=A_1+C/\tau,
$$
where $A_1$ and $C$ are nonzero $(t,\tau)$-independent matrices. Therefore, in this article we consider the family of models with
\begin{eqnarray}
\label{mero-def1}
 H&=&B_{00}t + B_{01}\tau +A_0,\\
\label{mero-def2}
H'&=&B_{11}\tau + B_{01} t +A_1 + C/\tau.
\end{eqnarray}
Substituting this pair into (\ref{cond1})-(\ref{cond2}), we find a set of conditions:
\begin{eqnarray}
\label{cc1}
&&[B_{00}, B_{11}] = [B_{00}, B_{01}]=[B_{11}, B_{01}]= 0,\\
\label{cc2}
&&[B_{01}, A_{0}] =[B_{00},A_{1}], \\
\label{cc6}
&&[B_{01},A_1]=[B_{11},A_0],\\
\label{cc3}
&&[B_{01}, C]  = -[A_{0}, A_{1}], \\
\label{cc4}
&&[B_{00}, C] = 0,\\
\label{cc5}
&&[A_{0}, C] = 0.
\end{eqnarray}
If $C=0$, then Eqs.~(\ref{cc1})-(\ref{cc3}) lead to the integrability conditions for the linear family (\ref{linear}), which we discussed in detail in \cite{large-class,multitime-LZ}.
Equation~(\ref{cc1}) tells that there is a basis, which we will call the {\it diabatic basis}, in which the matrices $B_{00}$, $B_{11}$, and $B_{10}$ are diagonal. 

Equation~(\ref{cc4}) shows then that off-diagonal elements of $C$ in this basis are possible only if the matrix $B_{00}$ has degenerate elements. Identifying $B_{00}$ with $B$ in (\ref{mlz}), this means that some of the diabatic energy levels in the MLZ model (\ref{mero-def1}) have to be parallel to each other, i.e. have the same slope in the time-energy plot. This distinguishes the class of solvable systems with nontrivial $C$ from the linear family (\ref{linear}). In contrast, the latter generally describes models with different slopes of all energy levels. 

Generally, Eqs.~(\ref{zero-curv-nonabel-2D-5a})-(\ref{zero-curv-nonabel-2D-5d}) allow more complex dependence of  $H'$ on $t$ and $\tau$. 
We will argue, however, that the pairs of operators (\ref{mero-def1}) and (\ref{mero-def2}) are of particular importance for the MLZ theory. Hence, we give a special name to this pair, {\it $t/\tau$-family},  for the property that all matrix components of $H'$ can be, maximum,  linear in $t$, whereas some of the components can also be inversely linear in $\tau$, and no other $\tau$-nonlinearity is allowed.

\section{Examples of nontrivial $t/\tau$-families}
It turns out that many known solvable MLZ Hamiltonians have a commuting operator of the form (\ref{mero-def2}). 

{\bf  I)  Driven Tavis-Cummings model}. This model describes interaction of $N$ spins with a single photonic mode that has a linearly  time-dependent frequency.  The Hamiltonian is
\be
\hat{H}_\mathrm{TC} \!=\! -t \hat{\psi}^{\dagger} \hat{\psi} +  \sum_{j=1}^{N} \left[ \tau \varepsilon_j \hat{s}_j^z +g \left(\hat{\psi}^{\dagger} \hat{s}^-_j + \hat{\psi} \hat{s}^{+}_j \right) \right],
\label{TC}
\ee
where $\hat{\psi}$ is the boson annihilation operator and $\hat{s}_j^z$, $\hat{s}_j^{\pm}$ are spin 1/2 operators. Note, that $N$ here is the number of spins rather than states. Hence, the model has a combinatorially large phase space for large $N$.

$\hat{H}_\mathrm{TC} $ is a member of a known family of linearly independent commuting operators \cite{commute,yuzbashyan-LZ}, one of which is
\begin{eqnarray}
\nonumber \hat{H}_\mathrm{TC}' \!=\!\sum_{j=1}^{N} \left[ \varepsilon_j (t +\tau \varepsilon_j) \hat{s}_j^z +g \varepsilon_j (\hat{\psi}^{\dagger} \hat{s}_j^- +\hat{\psi} \hat{s}_j^+) \right]+ \\
 +\frac{g^2}{\tau} \sum_{k,j, \, k\ne j} \hat{{\bf s}}_k \cdot \hat{{\bf s}}_j .
\label{htc-com1}
\ee

We can now read out the matrices that define the corresponding $t/\tau$-family:
\begin{eqnarray}
&& B_{00}=-\hat{\psi}^{\dagger} \hat{\psi}, \quad B_{01} =  \sum_{j=1}^{N} \varepsilon_j \hat{s}_j^z, \\
&& A_0=g \sum_{j=1}^{N} \left (\hat{\psi}^{\dagger} \hat{s}^-_j + \hat{\psi} \hat{s}^{+}_j \right) , \\
&& B_{11}=\sum_{j=1}^{N} \varepsilon_j ^2  \hat{s}_j^z, \quad A_1= g \sum_{j=1}^{N} \varepsilon_j (\hat{\psi}^{\dagger} \hat{s}_j^- +\hat{\psi} \hat{s}_j^+),\\
&& C= g^2\sum_{k,j, \, k\ne j} \hat{{\bf s}}_k \cdot \hat{{\bf s}}_j.
\label{tc-family}
\end{eqnarray}
The relations (\ref{cc1})-(\ref{cc5}) are straightforward to verify. 
For example, $B_{00}$, $B_{01}$ and $B_{11}$ are made of operators $\hat{\psi}^{\dagger} \hat{\psi}$ and $\hat{s}_j^z$ that commute with each other.  The least trivial to verify is the condition (\ref{cc5}), which is satisfied if  
\be
\left[ \sum_{j,k\ne j} \hat{{\bf s}}_k\cdot \hat{{\bf s}}_j, \sum_{r} \hat{s}^{+}_r \right]=0.
\label{comm1-mlz}
\ee
To prove this, we introduce operators $\hat{S}^{\pm}= \sum_{r=1}^N \hat{s}^{\pm}_r$, $\hat{S}_z= \sum_{r=1}^N \hat{s}^{z}_r$, and $\hat{S}^2=\hat{S}_z^2+\frac{1}{2}\left( \hat{S}^+\hat{S}^- +\hat{S}^-\hat{S}^+\right)$,
in terms of which 
\be
\sum_{j,k\ne j} \hat{{\bf s}}_k \cdot \hat{{\bf s}}_j = (\hat{S}^2/2-3N/4),
\label{comm2-mlz}
\ee
 where we used that $\hat{ \bf s}_j^2=1/2(1/2+1)=3/4$. $3N/4$ commutes with any operator, so we should only prove that 
\be
[\hat{S}^2, \hat{S}^{+} ]=0.
\label{comm3-mlz}
\ee 
Operators $\hat{S}_z$ and $\hat{S}^{\pm}$ satisfy standard spin algebra relations $[\hat{S}^+,\hat{S}^-]=2\hat{S}_z$, $[\hat{S}_z,\hat{S}^{\pm}]=\pm \hat{S}^{\pm}$. Hence, we have
\begin{eqnarray}
\nonumber &&[\hat{S}^2,\hat{S}^+] \!= \![\hat{S}_z^2,\hat{S}^+] \!+\! \frac{1}{2} [ \hat{S}^+\hat{S}^- +\hat{S}^-\hat{S}^+,\hat{S}^+]\! =\! \hat{S}_z[\hat{S}_z,\hat{S}^+] \!-\!  \\
&&-[\hat{S}^+,\hat{S}_z]\hat{S}_z + \frac{1}{2} \left(\hat{S}^+[\hat{S}^-,\hat{S}^+]+[\hat{S}^-,\hat{S}^+] \hat{S}^+ \right) = 0.
\label{proof-comm-mlz}
\end{eqnarray}

{\bf II) Interacting fermions}. Another solvable model with combinatorially complex phase space describes interaction of a single fermionic mode $\hat{d}$ with $(N-1)$ fermionic modes $\hat{c}_k$ \cite{quest-LZ}:
\be
\nonumber \hat{H}_{\rm F}\! =\! t \hd^{\dg} \hd \!+ \! \sum_{k=1}^{N-1} \Big[ \tau e_k (1\!-\!x \hd^{\dg} \hd) \hc_k^{\dg} \hc_k \!+ \!g_k (\hd^{\dg} \hc_k \!+\! \hc^{\dg}_k \hd) \Big],\\
 \label{Hf0}
\ee
where $e_k$, $x$, and $g_k$ are time-independent parameters.  For $x<1$, the transition probability matrix of this model coincides with the one for noninteracting fermions at $x=0$. For $x>1$, even the semiclassical description of this matrix is much more complex due to interference of many semiclassical trajectories \cite{quest-LZ}. 
Commuting operators for the Hamiltonian (\ref{Hf0}) were studied in \cite{yuzbashyan-LZ} but the set that was found there did not satisfy the first integrability condition  (\ref{cond1}) with $\hat{H}_{\rm F}$ .  
Here, we found that both $\hat{H}_{\rm F}$ and $\hat{H}_{\rm F}'$ can be constructed from the set of following commuting operators:
\be
&&\hat{H}_j({\bm e}) \!=\! t \hn_j(1-x \hn_d) -e_j \hn_j \!+\! x^2 \hn_d \hn_j \sum_{k} e_k \hn_k  \!  - \! \nn \\
&&~x  \hn_j \sum_k^{k\neq j} e_k \hn_k  \!-\!\  g_j (\hc^{\dg}_j \hd \!+\! \hd^{\dg} \hc_j) - x \hn_j \sum_k^{k\neq j} g_k (\hd^{\dg} \hc_k \!+\! \hc_k^{\dg} d)\!-\!  \nn \\
&& ~  \sum_k^{k\neq j} \frac{1 }{e_j -e_k}  \Big( g_k g_j( \hc_j^{\dg} \hc_k + \hc_k^{\dg} \hc_j ) - g_j^2 \hn_k -g_k^2 \hn_j \Big),
\ee
where $j=1, 2, \ldots, N-1$. Here, we introduce the number operators $\hat{n}_d \equiv \hd^{\dg} \hd$ and $\hn_j \equiv \hc^{\dg}_j \hc_j$. 
 Note that  $\hat{H}_j$ also satisfy the integrability condition (\ref{cond2}) when arbitrary two components of the vector ${\bm e}=(e_1,\ldots, e_{N-1})$ are taken as time variables. 

Commutativity of $\hat{H}_j$ with $\hat{H}_{\rm F}$ is easy to check at each power of $x$, and using simple identities, such as $\hat{n}_d \hat{n}_k \hat{d}^{\dagger} \hat{c}_k =0$. The longest calculations are then at zero power of $x$ but at $x=0$, operators $\hat{H}_{\rm F}$ and $\hat{H}_j$ are the secondary quantized versions  of the commuting operators for the Demkov-Osherov model \cite{patra-LZ}, whose integrability conditions (\ref{cond1}), (\ref{cond2}) were already verified in \cite{yuzbashyan-LZ}.

$\hat{H}_{\rm F}$ and $\hat{H}_j$ conserve the total number of particles $N_F$, i.e., $\hn_d + \sum_{j}^{N-1}\hn_j = N_F$. We have then for any  $N_F$ that
\be
\sum_{j} \hat{H}_j (\tau {\bm e}) \!= - \![1+ x(N_F-1)] \hat{H}_{\rm F} (t,\tau) +t N_F \hat{1},  
\label{Hf}
\ee
where  the last term is not important because it is proportional to the identity operator, and
\be
\hat{H}'_{\rm F} (t,\tau) = \sum_{j} e_j\hat{H}_j (\tau {\bm e}).
\label{Hprimef}
\ee
Since $\hat{H}_j (\tau {\bm e}) $ commute with each other for different $j$, the operators $\hat{H}_{\rm F}$ and $\hat{H}_{\rm F}'$ also commute with each other.   $\hat{H}_{\rm F}$ has the form (\ref{mero-def1}) and $\hat{H}'_{\rm F}$ has the form (\ref{mero-def2}). Hence, they make a  $t/\tau$-family. 

{\bf III) Demkov-Osherov and generalized bowtie models}. 
Two solvable MLZ models have been historically very influential. One is the Demkov-Osherov model \cite{do} in which the matrix $B_{00}$ has only one nonzero element. This model coincides with the single fermion sector of the previous fermionic model.   Hence the Demkov-Osherov model belongs to a $t/\tau$-family.

The generalized bowtie model describes interaction of two states, $|0_{+}\ra$ and $|0_{-}\ra$, which have parallel diabatic levels, with $N$ otherwise noninteracting states whose levels cross in one point \cite{gbow-tie}:
\be
\nonumber H_{bt} &=&  t\sum_{n=1}^N  \beta_n  |n\ra \la n| +\tau \left(|0_+\ra \la 0_+| - |0_{-} \ra \la 0_{-}| \right) +\\
 &&\sum_{n=1}^N g_n\left( |0_+\ra \la n| +|0_{-} \ra \la n| +{\rm h.c.} \right).
\label{bowtie1}
\ee
Commuting operators for this model have been explored in \cite{patra-LZ,yuzbashyan-LZ}, and one operator from the found set in \cite{yuzbashyan-LZ} has the form (\ref{mero-def2}). 
Let us show here how this operator could be derived from the conditions (\ref{cc1})-(\ref{cc5}).
In order to construct the corresponding $H'_{bt}$, we read the matrices $A_0$, $B_{00}$, and $B_{11}$ from (\ref{bowtie1}).
From (\ref{cc2}), it follows that 
\be
A_1 = \sum_{n=1}^N \frac{g_n}{\beta_n}  \left(|0_{-} \ra \la n| - |0_{+} \ra \la n| +{\rm h.c.} \right);
\label{a1-bt}
\ee
from (\ref{cc3})-(\ref{cc5}), we then find
\be
C\!=\! \kappa \left(|0_+\ra \la 0_+|\! +\! |0_{-} \ra \la 0_{-}|\! -\! |0_+\ra \la 0_{-}|\!- \!|0_{-} \ra \la 0_{+} | \right),
\label{c-bt}
\ee
where
$$
\kappa  =\sum_{n=1}^N g_n^2/\beta_n,
$$
and from (\ref{cc6}) we find 
\be
B_{11} = -\sum_{n=1}^{N} \frac{1}{\beta_n} |n\ra \la n|.
\label{b11-bt}
\ee
Thus, the bowtie Hamiltonian (\ref{bowtie1}) belongs to a nontrivial $t/\tau$-family with $H'_{bt} = B_{11}\tau +B_{10}t +A_1 +C/\tau$.


{\bf IV) Models generated by distortions}.
Finally, there are  known solvable models that were found by distortions of already solved models \cite{quest-LZ}, including distortions of the trivial models. The latter are the models that are either equivalent to several independent simpler Hamiltonians  that act in the direct product of their phase spaces simultaneously, or the models  that describe evolution of noninteracting identical fermions or bosons, so that the Heisenberg  equation for operators has the structure of some known solvable MLZ model \cite{yurovsky-triv,multiparticle}. 
The trivial models are used to obtain the connectivity graph that describes pairs of diabatic states that are directly coupled to each other. The distorted models are generated by assuming more general values of nonzero parameters, keeping the connectivity graph intact.  
One can then apply the integrability test \cite{quest-LZ} in order to derive new constraints on all these parameters. 

Such generated models are called distortions because they contain the previously solved models as special limits of some of the parameters.  It seems that any trivially solvable model can be distorted but usually such solvable MLZ models look somewhat artificial. Hence,
there are only three previously studied distorted MLZ models, namely two 6-state models and one 8-state model 
in  section~4 of Ref.~\cite{quest-LZ} and sections~7.2 and 7.3 of Ref.~\cite{large-class}. Similarly to the generalized bowtie model, we  verified (not shown) that all of them belong to some $t/\tau$-families, i.e., that their Hamiltonians have  nontrivial commuting operators of the form (\ref{mero-def2}).

 Given that the linear family \cite{multitime-LZ} is also a special case  of a pair (\ref{mero-def1}) and (\ref{mero-def2}) with $C=0$, 
we conclude that {\it almost  all} known solvable  MLZ Hamiltonians have  commuting operators of the form (\ref{mero-def2}). The only exceptions are found when all time-dependent diabatic levels cross in one point  but all such known  solvable models can also be derived and solved as special limits of some known solvable MLZ models whose diabatic levels cross in different points. Therefore, it is likely that all such finite size exceptional cases can, at least, be derived  from  the $t/\tau$-families.  
 
\section{Generating new solvable models from integrability conditions}

The integrability conditions (\ref{cc1})-(\ref{cc5}) 
are convenient for classifying solvable MLZ models for any given number $N$ of interacting states systematically. The approach is similar to the one that we described for the linear family in \cite{multitime-LZ}. 
The idea is to consider initially the general case of the matrices $B_{00}$, $B_{01}$ and $A_0$ for a given number of states $N$. 
In order to generate a $t/\tau$-family with nontrivial matrices $B_{01}$ and $C$, we should assume that $B_{00}$ has degenerate elements, so that
 $C$ has nonzero off-diagonal elements and $B_{01}$ is nondegenerate within this ``$B_{00}$-degenerate" subspace. The rest of $C$ must be diagonal.

\begin{figure}[!htb]
 \scalebox{0.27}[0.27]{\includegraphics{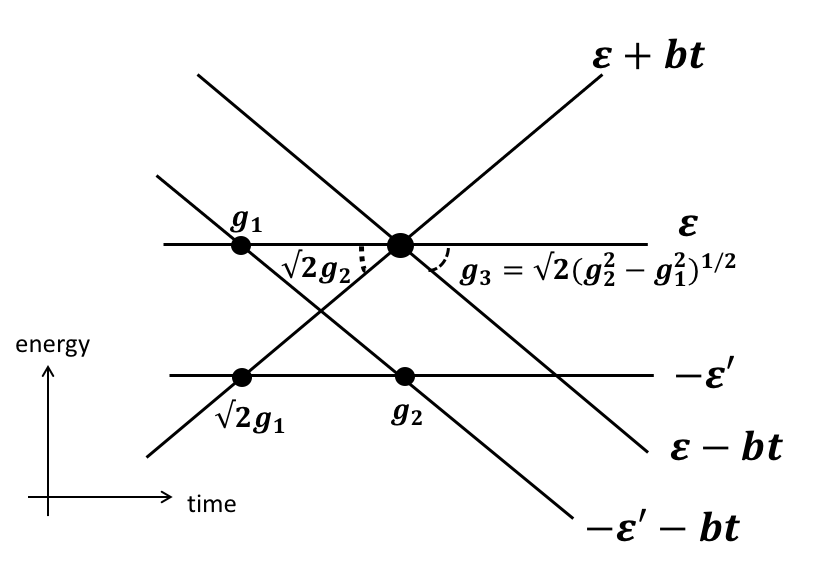}}
\caption{Time-dependence of the diabatic energy levels and the pairwise couplings in the integrable 5-state MLZ model (\ref{h51}) at $\tau=1$.}
\label{dld-5levels}
\end{figure}
\begin{figure}[!htb]
 \scalebox{0.27}[0.27]{\includegraphics{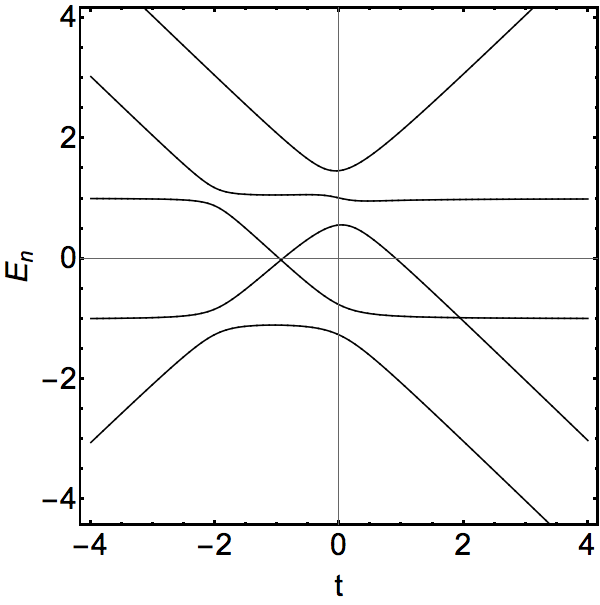}}
\caption{ Time-dependent eigenvalues of the Hamiltonian (\ref{h51}) for $e_1=1$, $e_2=1$, $g_1=0.15$, $g_2=0.25$, $b=1$, $\tau=1$.}
\label{spectrum-5levels}
\end{figure}
We can then treat conditions~(\ref{cc1})-(\ref{cc5})  as linear equations that determine elements of matrices $B_{11}$, $A_1$ and $C$. There are usually more equations in  (\ref{cc1})-(\ref{cc5})
 than parameters in these matrices. Therefore, after we eliminate all  parameters of $H'$, we end up with a set of generally nonlinear equations that constrain the parameters of the original MLZ model. 
 Very often these equations merely require that some of the couplings are zero but  some of the constraints can be more complex. Such restrictions often lead to physically uninteresting models that, e.g., have imaginary parameters
 or situations with not simply parallel but permanently degenerate diabatic levels. Sometimes, however, the constraints on the model parameters are neither degenerate nor unphysical.

By employing mathematical software, it is possible to investigate such cases with relatively small $N$ systematically. We did this for up to $N=6$ and also checked several connectivity graphs with $N=7,8$. Unfortunately, we found that most of these $t/\tau$-families  belonged to one of the already known cases from previous section. For example, we did not find new solvable MLZ models that described interactions of only two bands of parallel levels. Hence, it may happen that various sectors of the interacting fermion model with fixed numbers of fermions are the only possible solvable models with  two crossing bands.

Nevertheless, we did find several new solvable MLZ models of other types.  One of them describes interactions of five diabatic states with the Hamiltonian
\be
 \nonumber &&H_5 \! = \\
 &&\!\left(\begin{array}{ccccc}
e_1 \tau & 0 & g_1 & g_3 & g_2\sqrt{2} \\
0 & -e_2 \tau & g_2  & 0 & g_1 \sqrt{2} \\
g_1 & g_2 & -e_2 \tau -bt & 0 &0 \\
g_3 & 0 & 0 & e_1 \tau-bt & 0 \\
g_2\sqrt{2} & g_1 \sqrt{2} & 0& 0& e_1\tau+bt
\end{array}
 \right), 
\label{h51}
\ee 
where
$$
g_3=\sqrt{2(g_2^2-g_1^2)}, 
$$
and where $e_1$, $e_2$, $\tau$, $b$, $g_1$ and $g_2$ are free parameters. Time-dependence of the diabatic energy levels (diagonal elements of the matrix (\ref{h51})), and nonzero couplings for $H_5$ are illustrated in Fig.~\ref{dld-5levels}. This model describes interactions of  a single level and two bands, with two levels in each band. The time-dependent eigenvalues of the Hamiltonian of this model are given in Fig.~\ref{spectrum-5levels}, which shows that there are two points of exact pairwise level crossings. Such points are signatures of  integrable MLZ models \cite{quest-LZ}.

The corresponding commuting operator of the form (\ref{mero-def2}) is
\begin{widetext}
 \be
H_5'= \! \left(\begin{array}{ccccc}
e_1t +\frac{g_1^2}{b\tau} & -\frac{g_1g_2}{b\tau} & \frac{(e_1+e_2) g_1}{b} &0 &0 \\
\! -\frac{g_1g_2}{b\tau}\! & \frac{g_2^2}{b\tau} -\!\frac{e_2bt+(e_1+e_2)^2 \tau}{b}&0 & 0 & \frac{\sqrt{2}(e_1+e_2)g_1}{b} \\
\frac{(e_1+e_2) g_1}{b}  & 0 &  \frac{g_3^2}{2b\tau}  -\!\frac{e_2bt+(e_1+e_2)^2 \tau}{b} & -\frac{g_1g_3}{b\tau}  &0 \\
0 & 0 & -\frac{g_1g_3}{b\tau}  & e_1t+\frac{2g_1^2}{b\tau} & 0 \\
0 &  \frac{\sqrt{2}(e_1+e_2)g_1}{b}  & 0& 0& e_1t
\end{array}
\right).
\label{h52}
\ee 

Another example of a new solvable MLZ model, which we found following the same approach, describes interactions among six states and the spectrum that splits into bands with three, two, and  one levels: 
 \be
\nonumber H_6= \! \left(\begin{array}{cccccc}
e_1\tau &0&0&0&g&g\\
0&0&0&g&g&g\\
0&0& -e_2\tau &g&0&g\\
0&g&g&-e_2\tau+bt & 0& 0\\
g&g&0&0&e_1\tau+bt & 0 \\
g&g&g& 0&0& -bt
\end{array}
\right),
\label{h61}
\ee 
\end{widetext}
where $g$, $e_1$, $e_2$, $b$, $\tau$ are free parameters. 
In Fig.~\ref{spectrum-6levels}, we show the time-dependent eigenvalues of $H_6$ and mark the avoided crossing points by the strength of the corresponding level coupling. Note that there are four exact energy level crossing points that correspond to the crossings of diabatic levels that are not directly coupled to each other. This property of the model (\ref{h61}) is shared with all known solvable models whose diabatic levels experience only pairwise crossings \cite{quest-LZ}. 
\begin{figure}[!htb]
 \scalebox{0.27}[0.27]{\includegraphics{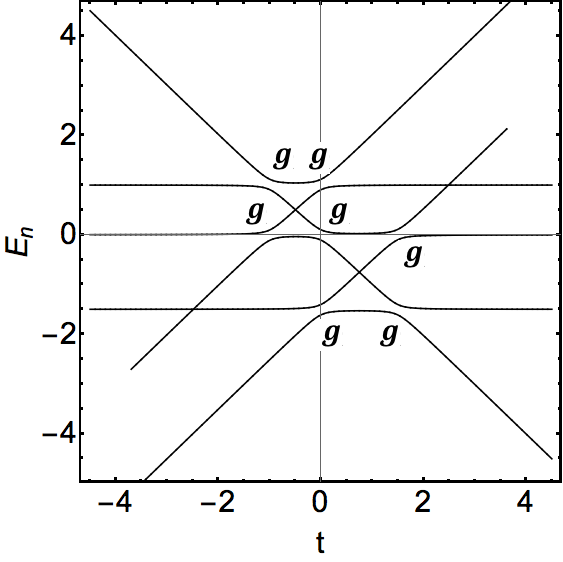}}
\caption{Time-dependent eigenvalues and nonzero level couplings of the Hamiltonian (\ref{h61}) for $e_1=1$, $e_2=1.5$, $g=0.105$, $b=1$, $\tau=1$.}
\label{spectrum-6levels}
\end{figure}

The matrices of transition probabilities for evolution from $t\rar-\infty$ to $t\rar +\infty$ can be constructed according to the rules from Ref.~\cite{quest-LZ}. For example, for the model in Fig.~\ref{spectrum-6levels}, there is no interference of different semiclassical trajectories, so the 
state-to-state transition probabilities can be read out from this figure. Let
$$
p\equiv e^{-2\pi g^2/\beta}, \quad q\equiv 1-p.
$$
The rules in \cite{quest-LZ} applied to (\ref{h61}) then produce
 \be
P_6=  \left(\begin{array}{cccccc}
 p^2 & q^2 & 0& 0&pq &pq\\
 pq^2   &p^3 & q^2&pq & p^2q &p^2q\\
 pq^2 & pq^2 & p^2& pq& q^3&p^2q\\
  q^3  &p^2q & pq &p^2 & pq^2&pq^2\\
  pq  & pq& 0& 0& p^2&q^2\\
 p^2q &p^2q& pq &q^2 &pq^2&p^3
\end{array}
\right).
\label{P61}
\ee 

We do not know  particular physical applications of the models (\ref{h51}) and (\ref{h61}). However, the very their existence is interesting because $H_5$ and $H_6$ cannot be found as  special cases of larger known models or as a distortion of some already known solvable model. All distorted  models with up to six interacting states
have been already studied in \cite{quest-LZ} and \cite{large-class}. They do not contain five-state systems, and they have different connectivity graphs from the one for $H_6$. This property, as well as the simplicity of the Hamiltonians $H_5$ and $H_6$, indicate that such systems can be low-$N$ instances of some, still unknown, nontrivial model of interacting spins, fermions, or bosons, akin to the interacting fermions or driven Tavis-Cummings  models. To verify this conjecture, we should explore other realizations of the algebra (\ref{cc1})-(\ref{cc5}), and additional studies of systems with large $N$ would be useful. We leave this direction for the future studies.

Finally, we note that in addition to  the pairs of real symmetric Hamiltonians  that had the form (\ref{mero-def1})-(\ref{mero-def2}), we observed many nontrivial pairs of this type that contained imaginary parameters. 
Although such pairs of operators cannot describe unitary evolution of a finite number of states, such operators do describe unitary evolution if corresponding equations are interpreted as Heisenberg equations for bosonic operators.
Namely, if the number of bosons is not conserved but the secondary quantized Hamiltonian is quadratic in bosonic operators, then the Heisenberg equations  have the form of a non-unitary evolution for amplitudes of a finite size quantum system.
 Hence, there is a large class of solvable MLZ models that describes interacting bosonic systems without particle conservation. 
Such MLZ systems have been studied for applications to dynamic transitions through the Feshbach resonance in ultracold atoms \cite{yurovski,altland-LZ,chains-LZ} but their classification remains essentially unexplored.

\section{Properties of MLZ models from $t/\tau$-family}

MLZ models that belong to $t/\tau$-families have many features that are common with the linear family (\ref{linear}) and can be derived analogously. The linear family was studied in detail \cite{large-class,multitime-LZ}. Therefore, here we provide only brief discussion of these features. For further information, we refer to Ref.~\cite{quest-LZ} and section~8 in Ref.~\cite{large-class}.

Following \cite{commute}, let us define the evolution operator for a path ${\cal P}$ in $(t,\tau)$-space:
\be
\nonumber U=\hat{\cal T} {\rm exp}\left[-i\int_{\cal{P}} (H\, d t +H'\, d\tau) \right],
\label{path1}
\ee
where $\hat{\cal T}$ is the path ordering operator.
The scattering matrices for MLZ models can be constructed by deforming the physical evolution path ${\cal P}$ in $(t,\tau)$-space so that the deformed path ${\cal P}_{\infty}$  encloses the region with complex nonadiabatic transitions from a large distance, as shown in Fig.~\ref{pathes-pic}. Note that the parameter $\tau$ merely defines the scale of the diabatic level splitting in the bands of $H$. Hence, without loss of generality, we can assume  that $\tau=1$ corresponds to the physical evolution, for which we should find the evolution matrix for time interval $t\in (-T,T)$, where $T \rar \infty$. 
\begin{figure}[!htb]
 \scalebox{0.3}[0.3]{\includegraphics{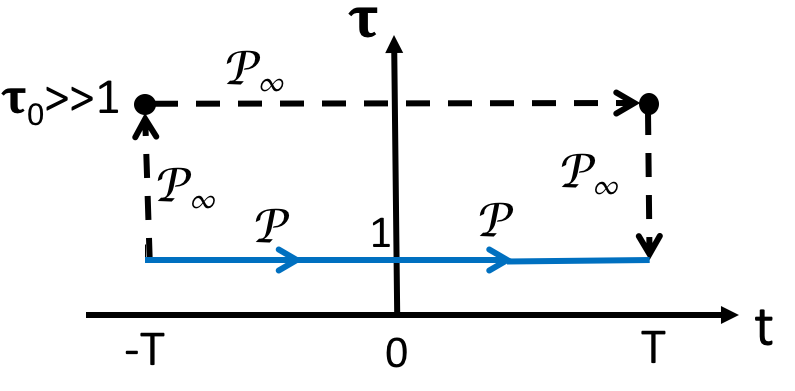}}
\caption{Evolution path ${\cal P}$ (solid blue arrow) with MLZ  Hamiltonian, at $\tau=1$ and for $T\rar \infty$, can be deformed into ${\cal P}_{\infty}$ such that the evolution along the horizontal part of ${\cal P}_{\infty}$ has the same MLZ Hamiltonian but at  $\tau=\tau_0 \gg 1$.  }
\label{pathes-pic}
\end{figure}

Integrability conditions (\ref{cond1}) and (\ref{cond2}) guarantee that we can deform ${\cal P}$ into ${\cal P}_{\infty}$ without changing the final evolution matrix, as long as the initial and the final points of the paths are the same \cite{commute,faddeev-book,deform-book}. 
During such a path deformation, a certain parameter combination becomes a new time variable. This idea has been used before in  quantum field theory \cite{kirzhnitz}. 
Specifically, for $T\rar \infty$, it is convenient to choose 
${\cal P}_{\infty}$ with the two vertical legs that connect points at $\tau=1$ and at $\tau=\tau_0 \gg 1$. Due to large $T$, the evolution along these legs is purely adiabatic. Hence, the vertical legs contribute only to the phases of scattering amplitudes but not to the transition probabilities. The remaining part of ${\cal P}_{\infty}$ is horizontal in $(t,\tau)$-plane. Hence, it is described by the MLZ Hamiltonian $H$  at $\tau=\tau_0 \gg 1$. At large $\tau$, nonadiabatic transitions between pairs of diabatic states become well separated  
in time, that is along the horizontal part of ${\cal P}_{\infty}$ in Fig.~\ref{pathes-pic}. This means that we can sequentially apply the Landau-Zener formula for transitions near each diabatic level crossing point. Then, the final scattering matrix is  the product of the scattering matrices for pairwise 
level crossings and the trivial diagonal matrices that describe intervals with adiabatic evolution.  This logic applies equally well to the linear family \cite{large-class} and the more general $t/\tau$-family. The only difference is that  $\sim 1/\tau$ terms in $H'$  contribute to the adiabatic phase along ${\cal P}_{\infty}$, which is easy to calculate for $t/\tau$-family analytically.   

Two  properties of solvable MLZ models were used in our early articles as a test for integrability \cite{quest-LZ}. The first one is the ``zero area property". It means that any loop in the diabatic level diagram encloses zero area if clockwise and counterclockwise enclosed regions are counted with opposite signs. For $t/\tau$-family, this rule follows from Eqs.~(\ref{cc2}) and (\ref{cc6}). Taking the   matrix component in both sides of this equation for diabatic level indices $a$ and $b$, we find two linear equations on $A_{0}^{ab}$ and $A_{1}^{ab}$ that have a nonzero solution if 
\be
B_{11}^{aa} -B_{11}^{bb} =\frac{(B_{01}^{aa} -B_{01}^{bb})^2}{B_{00}^{aa}-B_{00}^{bb}}.
\label{zerosum1}
\ee
Consider all indices $i_1,\ldots,i_{N(L)}$ of diabatic levels that encounter sequentially along some loop $L$ on the connectivity graph of the model. Links of this graph correspond to nonzero direct off-diagonal couplings in $H$. Summing expressions (\ref{zerosum1}) for all nodes of this loop we find 
\be
\sum_{k,\,i_k\in L} \frac{(B_{01}^{i_k i_k} -B_{01}^{i_{k+1} i_{k+1}})^2}{B_{00}^{i_k i_k}-B_{00}^{i_{k+1} i_{k+1}}}=0.
\label{zerosum2}
\ee
According to section~8.7 of Ref.~\cite{large-class}, this is precisely the zero area condition for the MLZ model (\ref{mero-def1}). 

The second property of solvable MLZ models is the existence of a special number of exact crossing points in their time-dependent spectra \cite{quest-LZ}.
Exact eigenvalue level crossing points in MLZ Hamiltonians have previously been discussed in many contexts \cite{bow-tie1,wernsdorfer,nagaosa-LZ,multiparticle,patra-LZ,li-kramers}. However, in the currently known fully solvable MLZ models,  there must be precisely one crossing point per every isolated pair of crossing diabatic levels, whose diabatic states are not coupled to each other directly. Such crossing points must appear at sufficiently small but finite values of all off-diagonal couplings but can annihilate with each other at large coupling values \cite{quest-LZ}. This number is different from, e.g., the number of exact crossing points in the  MLZ Hamiltonians that were generated as commuting partners of the Demkov-Osherov Hamiltonian \cite{patra-LZ} (when $t$ is still interpreted as time there). Hence, the latter models are  generally not solvable MLZ models in the sense that one cannot write their scattering matrices explicitly as elementary matrix products defined in \cite{quest-LZ}, and analytical forms of their scattering matrices remain unknown.

The presence of specific exact crossing points in the Hamiltonian $H$ from  a $t/\tau$-family can be explained analogously to how this was done for the linear families in section~8.8 of Ref.~\cite{large-class}. Namely, by rescaling time, the ``sufficiently small coupling" condition is transformed to the condition that the exact crossing must appear along the path ${\cal P}_{\infty}$ for sufficiently large but finite $T$ and $\tau_0$. 

Then, the two diabatic levels of $H$ with different slopes and zero direct coupling between them must cross at some point on  ${\cal P}_{\infty}$. Imagine that there is no corresponding exact crossing of two Hamiltonian eigenvalues in the vicinity of this point. Then, the corresponding diabatic states are coupled in high-order perturbation series over $1/\tau_0$ at this point. Hence, the eigenstate of $H$ at this point must be in a nontrivial superposition of the two corresponding diabatic states.

However, the corresponding diagonal elements of $H'$ at the same point remain non-degenerate.  Indeed, Eq.~(\ref{zerosum1}) is the only condition that relates slopes of diabatic levels $a$ and $b$, and this condition applies only to the case with $A_0^{ab} \ne 0$. 
If $A_0^{ab} = 0$ then we generally should assume that slopes of the corresponding diabatic levels $a$ and $b$ in $B_{00}$ and $B_{11}$ are not related by Eq.~(\ref{zerosum1}). 
Moreover, in $H'$ the corresponding diabatic energy level splitting is large at the same point  of ${\cal P}_{\infty}$
due to large values of  $\tau_0$. 

This means that the corresponding eigenstates of $H'$ coincide with diabatic states up to corrections that vanish as $T,\tau_0 \rar \infty$. 
On the other hand, $H$ and $H'$ commute. Their eigenstates coincide and, therefore, cannot be in a nontrivial superposition. We arrive at contradiction, meaning that along the path ${\cal P}_{\infty}$ for sufficiently large $T$ and $\tau_0$ in Fig.~\ref{pathes-pic} there will be exact energy level crossings that correspond to pairs of diabatic levels with zero direct couplings between the corresponding diabatic states.

\section{Discussion}

The present article essentially completes the series of our papers in the journal, J. Phys. A: Math. Theor. \cite{quest-LZ,large-class,multitime-LZ}, about solvable MLZ systems. We found the minimal symmetry that is sufficient to explain the existence of all such known models with finite phase space.
This symmetry is the existence  of a nontrivial Hermitian operator (\ref{mero-def2}) whose elements satisfy conditions (\ref{cc1})-(\ref{cc5}). All known facts about construction of MLZ model's solution and about the integrability conditions on the parameters, which were described in \cite{quest-LZ} as conjectures, are now proved and very well understood. 

This does not mean that there are no more open questions about integrability in MLZ  theory; just the new questions are beyond the story that we started in \cite{quest-LZ}. For example, conditions (\ref{zero-curv-nonabel-2D-5a})-(\ref{zero-curv-nonabel-2D-5d})  definitely  have other families as solutions. In fact, the simple bowtie model belongs to such a family \cite{yuzbashyan-LZ} but this model can be derived and solved as a specific limit of the generalized bowtie model \cite{bow-tie1} that belongs to a $t/\tau$-family. Hence, the question about existence of essentially different MLZ models, whose scattering matrices can be found analytically, remains open. If integrable families leading to such models will be found, then unusual physical phenomena can be revealed. 
The only known case, however, that does not belong to a known integrable family is the infinite linear Landau-Zener chain with nonzero other than nearest neighbor couplings \cite{lz-chain}. This Hamiltonian  has unbound spectrum, so it may be the only physically interesting exception. 

It is also expected that there are integrable families of the form (\ref{cond1}) and (\ref{cond2}), with one of the Hamiltonians having the form (\ref{mlz}), but such that it is insufficient to find the explicit solution of this MLZ problem. Hence, another open question is how such symmetries can assist our understanding of complex explicitly time-dependent physics beyond providing the explicit scattering matrices.

Apart from solving time-dependent Schr\"odinger equation, our studies contributed to understanding of another long-standing problem, namely the emergence of exact crossing points in integrable quantum models at seemingly unpredictable values of parameters \cite{shastry1}. Our theory predicts the precise number of exact energy level crossing points in solvable MLZ Hamiltonians at small values of couplings, and explains why this number does not generally have to be conserved at large couplings.  
We would also like to mention  another application of solvable MLZ models. Their formal perturbative solutions are expressed via repeated multiple Gaussian integrals over a nontrivial domain of variables.  Hence, solvable MLZ models have helped to develop a method to unbundle such complex integrals and take them explicitly \cite{fai}.

Our examination of the few-state $t/\tau$-families revealed two cases, with five and six state phase space, that did not belong to any previously known model or its distortion. This can mean that there are still unidentified integrable Hamiltonians that may describe useful models of interacting bosons and fermions in linearly time-dependent fields. The previously identified such systems, namely, interacting fermions \cite{quest-LZ}, $\gamma$-magnets \cite{multitime-LZ}, and driven Tavis-Cummings \cite{commute} models revealed unusual effects, such as dynamic spin localization and dynamic phase transitions. Therefore, identification of other solvable models with comparable complexity can be beneficial for theoretical physics. A possible path forward in this direction is to use observation in \cite{yuzbashyan-LZ} that several solvable MLZ systems can be obtained after  changes of variables in the family of Gaudin magnet Hamiltonians. So far, this observation has not produced new solvable MLZ systems but one can try to reproduce our newfound systems from Gaudin magnets and then generalizations may follow.

Explicitly time-dependent quantum problems with physically interesting and nonperturbative time-dependence of parameters are usually considered unapproachable by analytical methods. There are many reasons for this: no energy conservation, generation of entanglement;  such problems do not reduce to identification of the properties of the ground state or matrix diagonalization. In our series of articles we demonstrated that this perception is misleading. 

The MLZ Hamiltonians of the form $H=A+Bt$ are found very generally in quantum hysteresis \cite{wernsdorfer}, dynamic passage through Feshbach resonance \cite{altland}, and quantum annealing \cite{dzyarmaga}. We showed that solvable models of this type are actually numerous and they produce simple explicit analytical solutions that describe nontrivial many-body dynamics. Hence, we hope that our results will stimulate studies of other explicitly time-dependent problems. For example, it is straightforward to generalize our approach to the Hamiltonians of the form $H=A+Bt+C/t$ (see e.g., \cite{yuzbashyan-LZ,LZC,bcs,inverse1,inverse2}).

\section*{Acknowledgements}
This work was supported by the U.S. Department of Energy, Office of Science, Basic Energy
Sciences, Materials Sciences and Engineering Division, Condensed Matter Theory Program (V.Y. C. and N.A.S.), and by the J. Michael Kosterlitz Postdoctoral Fellowship at Brown University (C.S.). F. Li was supported by NSFC (No. 11905054) and by the Fundamental Research Funds for the Central Universities from China.   


\end{document}